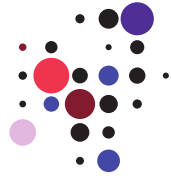

# The IceCube Neutrino Observatory, the Pierre Auger Observatory and the Telescope Array: Joint Contribution to the 34th International Cosmic Ray Conference (ICRC 2015)

## The IceCube Collaboration

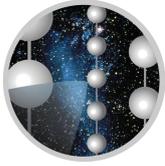

M.G. Aartsen[2], K. Abraham[32], M. Ackermann[48], J. Adams[15], J.A. Aguilar[12], M. Ahlers[29], M. Ahrens[39], D. Altmann[23], T. Anderson[45], I. Ansseau[12], M. Archinger[30], C. Arguelles[29], T.C. Arlen[45], J. Auffenberg[1], X. Bai[37], S.W. Barwick[26], V. Baum[30], R. Bay[7], J.J. Beatty[17,18], J. Becker Tjus[10], K.-H. Becker[47], E. Beiser[29], S. BenZvi[29], P. Berghaus[48], D. Berley[16], E. Bernardini[48], A. Bernhard[32], D.Z. Besson[27], G. Binder[8,7], D. Bindig[47], M. Bissok[1], E. Blaufuss[16], J. Blumenthal[1], D.J. Boersma[46], C. Bohm[39], M. Börner[20], F. Bos[10], D. Bose[41], S. Böser[30], O. Botner[46], J. Braun[29], L. Brayeur[13], H.-P. Bretz[48], N. Buzinsky[22], J. Casey[5], M. Casier[13], E. Cheung[16], D. Chirkin[29], A. Christov[24], K. Clark[42], L. Classen[23], S. Coenders[32], D.F. Cowen[45,44], A.H. Cruz Silva[48], J. Daughhetee[5], J.C. Davis[17], M. Day[29], J.P.A.M. de André[21], C. De Clercq[13], E. del Pino Rosendo[30], H. Dembinski[33], S. De Ridder[25], P. Desiati[29], K.D. de Vries[13], G. de Wasseige[13], M. de With[9], T. DeYoung[21], J.C. Díaz-Vélez[29], V. di Lorenzo[30], J.P. Dumm[39], M. Dunkman[45], R. Eagan[45], B. Eberhardt[30], T. Ehrhardt[30], B. Eichmann[10], S. Euler[46], P.A. Evenson[33], O. Fadiran[29], S. Fahey[29], A.R. Fazely[6], A. Fedynitch[10], J. Feintzeig[29], J. Felde[16], K. Filimonov[7], C. Finley[39], T. Fischer-Wasels[47], S. Flis[39], C.-C. Fösig[30], T. Fuchs[20], T.K. Gaisser[33], R. Gaior[14], J. Gallagher[28], L. Gerhardt[8,7], K. Ghorbani[29], D. Gier[1], L. Gladstone[29], M. Glagla[1], T. Glüsenkamp[48], A. Goldschmidt[8], G. Golup[13], J.G. Gonzalez[33], D. Góra[48], D. Grant[22], J.C. Groh[45], A. Groß[32], C. Ha[8,7], C. Haack[1], A. Haj Ismail[25], A. Hallgren[46], F. Halzen[29], B. Hansmann[1], K. Hanson[29], D. Hebecker[9], D. Heereman[12], K. Helbing[47], R. Hellauer[16], D. Hellwig[1], S. Hickford[47], J. Hignight[21], G.C. Hill[2], K.D. Hoffman[16], R. Hoffmann[47], K. Holzapfel[32], A. Homeier[11], K. Hoshina[29,a], F. Huang[45], M. Huber[32], W. Huelsnitz[16], P.O. Hulth[39], K. Hultqvist[39], S. In[41], A. Ishihara[14], E. Jacobi[48], G.S. Japaridze[4], K. Jero[29], M. Jurkovic[32], B. Kaminsky[48], A. Kappes[23], T. Karg[48], A. Karle[29], M. Kauer[29,34], A. Keivani[45], J.L. Kelley[29], J. Kemp[1], A. Kheirandish[29], J. Kiryluk[40], J. Kläs[47], S.R. Klein[8,7], G. Kohnen[31], R. Koirala[33], H. Kolanoski[9], R. Konietz[1], A. Koob[1], L. Köpke[30], C. Kopper[22], S. Kopper[47], D.J. Koskinen[19], M. Kowalski[9,48], K. Krings[32], G. Kroll[30], M. Kroll[10], J. Kunnen[13], N. Kurahashi[36], T. Kuwabara[14], M. Labare[25], J.L. Lanfranchi[45], M.J. Larson[19], M. Lesiak-Bzdak[40], M. Leuermann[1], J. Leuner[1], L. Lu[14], J. Lünemann[13], J. Madsen[38], G. Maggi[13], K.B.M. Mahn[21], R. Maruyama[34], K. Mase[14], H.S. Matis[8], R. Maunu[16], F. McNally[29], K. Meagher[12], M. Medici[19], A. Meli[25], T. Menne[20], G. Merino[29], T. Meures[12], S. Miarecki[8,7], E. Middell[48], E. Middlemas[29], L. Mohrmann[48], T. Montaruli[24], R. Morse[29], R. Nahnhauer[48], U. Naumann[47], G. Neer[21], H. Niederhausen[40], S.C. Nowicki[22], D.R. Nygren[8], A. Obertacke[47], A. Olivas[16], A. Omairat[47], A. O'Murchadha[12], T. Palczewski[43], H. Pandya[33], L. Paul[1], J.A. Pepper[43], C. Pérez de los Heros[46], C. Pfendner[17], D. Pieloth[20], E. Pinat[12], J. Posselt[47], P.B. Price[7], G.T. Przybylski[8], J. Pütz[1], M. Quinnan[45], C. Raab[12], L. Rädel[1], M. Rameez[24], K. Rawlins[3], R. Reimann[1], M. Relich[14], E. Resconi[32], W. Rhode[20], M. Richman[36], S. Richter[29], B. Riedel[22], S. Robertson[2], M. Rongen[1], C. Rott[41], T. Ruhe[20], D. Ryckbosch[25], S.M. Saba[10], L. Sabbatini[29], H.-G. Sander[30], A. Sandrock[20], J. Sandroos[30], S. Sarkar[19,35], K. Schatto[30], F. Scheriau[20], M. Schimp[1], T. Schmidt[16], M. Schmitz[20], S. Schoenen[1], S. Schöneberg[10], A. Schönwald[48], L. Schulte[11], D. Seckel[33], S. Seunarine[38], R. Shanidze[48], M.W.E. Smith[45], D. Soldin[47], M. Song[16], G.M. Spiczak[38], C. Spiering[48],

M. Stahlberg[1], M. Stamatikos[17,b], T. Stanev[33], N.A. Stanisha[45], A. Stasik[48], T. Stezelberger[8], R.G. Stokstad[8], A. Stößl[48], R. Ström[46], N.L. Strotjohann[48], G. W. Sullivan[16], M. Sutherland[17], H. Taavola[46], I. Taboada[5], S. Ter-Antonyan[6], A. Terliuk[48], G. Tešić[45], S. Tilav[33], P.A. Toale[43], M.N. Tobin[29], S. Toscano[13], D. Tosi[29], M. Tselengidou[23], A. Turcati[32], E. Unger[46], M. Usner[48], S. Vallecorsa[24], J. Vandenbroucke[29], N. van Eijndhoven[13], S. Vanheule[25], J. van Santen[29], J. Veenkamp[32], M. Vehring[1], M. Voge[11], M. Vraeghe[25], C. Walck[39], A. Wallace[2], M. Wallraff[1], N. Wandkowsky[29], Ch. Weaver[22], C. Wendt[29], S. Westerhoff[29], B.J. Whelan[2], N. Whitehorn[29], C. Wichary[1], K. Wiebe[30], C.H. Wiebusch[1], L. Wille[29], D.R. Williams[43], H. Wissing[16], M. Wolf[39], T.R. Wood[22], K. Woschnagg[7], D.L. Xu[43], X.W. Xu[6], Y. Xu[40], J.P. Yanez[48], G. Yodh[26], S. Yoshida[14], M. Zoll[39]

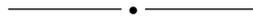

[1] III. Physikalisches Institut, RWTH Aachen University, D-52056 Aachen, Germany
[2] School of Chemistry & Physics, University of Adelaide, Adelaide SA, 5005 Australia
[3] Dept. of Physics and Astronomy, University of Alaska Anchorage, 3211 Providence Dr., Anchorage, AK 99508, USA
[4] CTSPS, Clark-Atlanta University, Atlanta, GA 30314, USA
[5] School of Physics and Center for Relativistic Astrophysics, Georgia Institute of Technology, Atlanta, GA 30332, USA
[6] Dept. of Physics, Southern University, Baton Rouge, LA 70813, USA
[7] Dept. of Physics, University of California, Berkeley, CA 94720, USA
[8] Lawrence Berkeley National Laboratory, Berkeley, CA 94720, USA
[9] Institut für Physik, Humboldt-Universität zu Berlin, D-12489 Berlin, Germany
[10] Fakultät für Physik & Astronomie, Ruhr-Universität Bochum, D-44780 Bochum, Germany
[11] Physikalisches Institut, Universität Bonn, Nussallee 12, D-53115 Bonn, Germany
[12] Université Libre de Bruxelles, Science Faculty CP230, B-1050 Brussels, Belgium
[13] Vrije Universiteit Brussel, Dienst ELEM, B-1050 Brussel, Belgium
[14] Dept. of Physics, Chiba University, Chiba 263-8522, Japan
[15] Dept. of Physics and Astronomy, University of Canterbury, Private Bag 4800, Christchurch, New Zealand
[16] Dept. of Physics, University of Maryland, College Park, MD 20742, USA
[17] Dept. of Physics and Center for Cosmology and Astro-Particle Physics, Ohio State University, Columbus, OH 43210, USA
[18] Dept. of Astronomy, Ohio State University, Columbus, OH 43210, USA
[19] Niels Bohr Institute, University of Copenhagen, DK-2100 Copenhagen, Denmark
[20] Dept. of Physics, TU Dortmund University, D-44221 Dortmund, Germany
[21] Dept. of Physics and Astronomy, Michigan State University, East Lansing, MI 48824, USA
[22] Dept. of Physics, University of Alberta, Edmonton, Alberta, Canada T6G 2E1
[23] Erlangen Centre for Astroparticle Physics, Friedrich-Alexander-Universität Erlangen-Nürnberg, D-91058 Erlangen, Germany
[24] Département de physique nucléaire et corpusculaire, Université de Genève, CH-1211 Genève, Switzerland
[25] Dept. of Physics and Astronomy, University of Gent, B-9000 Gent, Belgium
[26] Dept. of Physics and Astronomy, University of California, Irvine, CA 92697, USA
[27] Dept. of Physics and Astronomy, University of Kansas, Lawrence, KS 66045, USA
[28] Dept. of Astronomy, University of Wisconsin, Madison, WI 53706, USA
[29] Dept. of Physics and Wisconsin IceCube Particle Astrophysics Center, University of Wisconsin, Madison, WI 53706, USA
[30] Institute of Physics, University of Mainz, Staudinger Weg 7, D-55099 Mainz, Germany
[31] Université de Mons, 7000 Mons, Belgium
[32] Technische Universität München, D-85748 Garching, Germany

[33] Bartol Research Institute and Dept. of Physics and Astronomy, University of Delaware, Newark, DE 19716, USA
[34] Dept. of Physics, Yale University, New Haven, CT 06520, USA
[35] Dept. of Physics, University of Oxford, 1 Keble Road, Oxford OX1 3NP, UK
[36] Dept. of Physics, Drexel University, 3141 Chestnut Street, Philadelphia, PA 19104, USA
[37] Physics Department, South Dakota School of Mines and Technology, Rapid City, SD 57701, USA
[38] Dept. of Physics, University of Wisconsin, River Falls, WI 54022, USA
[39] Oskar Klein Centre and Dept. of Physics, Stockholm University, SE-10691 Stockholm, Sweden
[40] Dept. of Physics and Astronomy, Stony Brook University, Stony Brook, NY 11794-3800, USA
[41] Dept. of Physics, Sungkyunkwan University, Suwon 440-746, Korea
[42] Dept. of Physics, University of Toronto, Toronto, Ontario, Canada, M5S 1A7
[43] Dept. of Physics and Astronomy, University of Alabama, Tuscaloosa, AL 35487, USA
[44] Dept. of Astronomy and Astrophysics, Pennsylvania State University, University Park, PA 16802, USA
[45] Dept. of Physics, Pennsylvania State University, University Park, PA 16802, USA
[46] Dept. of Physics and Astronomy, Uppsala University, Box 516, S-75120 Uppsala, Sweden
[47] Dept. of Physics, University of Wuppertal, D-42119 Wuppertal, Germany
[48] DESY, D-15735 Zeuthen, Germany
[a] Earthquake Research Institute, University of Tokyo, Bunkyo, Tokyo 113-0032, Japan
[b] NASA Goddard Space Flight Center, Greenbelt, MD 20771, USA

# The Pierre Auger Collaboration

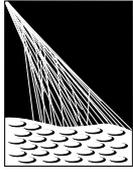

A. Aab[41], P. Abreu[65], M. Aglietta[52], E.J. Ahn[80], I. Al Samarai[28], I.F.M. Albuquerque[16], I. Allekotte[1], P. Allison[85], A. Almela[11,8], J. Alvarez Castillo[58], J. Alvarez-Muñiz[75], R. Alves Batista[40], M. Ambrosio[43], A. Aminaei[59], G.A. Anastasi[45], L. Anchordoqui[79], S. Andringa[65], C. Aramo[43], F. Arqueros[72], N. Arsene[68], H. Asorey[1,24], P. Assis[65], J. Aublin[30], G. Avila[10], N. Awal[83], A.M. Badescu[69], C. Baus[35], J.J. Beatty[85], K.H. Becker[34], J.A. Bellido[12], C. Berat[31], M.E. Bertaina[52], X. Bertou[1], P.L. Biermann[38], P. Billoir[30], S.G. Blaess[12], A. Blanco[65], M. Blanco[30], J. Blazek[26], C. Bleve[47], H. Blümer[35,36], M. Boháčová[26], D. Boncioli[51], C. Bonifazi[22], N. Borodai[63], J. Brack[78], I. Brancus[66], T. Bretz[39], A. Bridgeman[36], P. Brogueira[65], P. Buchholz[41], A. Bueno[74], S. Buitink[59], M. Buscemi[43], K.S. Caballero-Mora[56], B. Caccianiga[42], L. Caccianiga[30], M. Candusso[44], L. Caramete[67], R. Caruso[45], A. Castellina[52], G. Cataldi[47], L. Cazon[65], R. Cester[46], A.G. Chavez[57], A. Chiavassa[52], J.A. Chinellato[17], J. Chudoba[26], M. Cilmo[43], R.W. Clay[12], G. Cocciolo[47], R. Colalillo[43], A. Coleman[86], L. Collica[42], M.R. Coluccia[47], R. Conceição[65], F. Contreras[9], M.J. Cooper[12], A. Cordier[29], S. Coutu[86], C.E. Covault[76], R. Dallier[33,32], B. Daniel[17], S. Dasso[5,3], K. Daumiller[36], B.R. Dawson[12], R.M. de Almeida[23], S.J. de Jong[59,61], G. De Mauro[59], J.R.T. de Mello Neto[22], I. De Mitri[47], J. de Oliveira[23], V. de Souza[15], L. del Peral[73], O. Deligny[28], N. Dhital[82], C. Di Giulio[44], A. Di Matteo[48], J.C. Diaz[82], M.L. Díaz Castro[17], F. Diogo[65], C. Dobrigkeit[17], W. Docters[60], J.C. D'Olivo[58], A. Dorofeev[78], Q. Dorosti Hasankiadeh[36], R.C. dos Anjos[15], M.T. Dova[4], J. Ebr[26], R. Engel[36], M. Erdmann[39], M. Erfani[41], C.O. Escobar[80,17], J. Espadanal[65], A. Etchegoyen[8,11], H. Falcke[59,62,61], K. Fang[87], G. Farrar[83], A.C. Fauth[17], N. Fazzini[80], A.P. Ferguson[76], B. Fick[82], J.M. Figueira[8], A. Filevich[8], A. Filipčič[70,71], O. Fratu[69], M.M. Freire[6], T. Fujii[87], B. García[7], D. García-Gámez[29], D. Garcia-Pinto[72], F. Gate[33], H. Gemmeke[37], A. Gherghel-Lascu[66], P.L. Ghia[30], U. Giaccari[22], M. Giammarchi[42], M. Giller[64], D. Głas[64], C. Glaser[39], H. Glass[80], G. Golup[1], M. Gómez Berisso[1], P.F. Gómez Vitale[10], N. González[8], B. Gookin[78], J. Gordon[85], A. Gorgi[52], P. Gorham[88], P. Gouffon[16], N. Griffith[85], A.F. Grillo[51], T.D. Grubb[12], F. Guarino[43], G.P. Guedes[18], M.R. Hampel[8], P. Hansen[4], D. Harari[1], T.A. Harrison[12], S. Hartmann[39], J.L. Harton[78], A. Haungs[36], T. Hebbeker[39], D. Heck[36], P. Heimann[41], A.E. Hervé[36], G.C. Hill[12], C. Hojvat[80], N. Hollon[87], E. Holt[36], P. Homola[34], J.R. Hörandel[59,61], P. Horvath[27], M. Hrabovský[27,26], D. Huber[35], T. Huege[36], A. Insolia[45], P.G. Isar[67], I. Jandt[34], S. Jansen[59,61], C. Jarne[4], J.A. Johnsen[77], M. Josebachuili[8], A. Kääpä[34], O. Kambeitz[35], K.H. Kampert[34], P. Kasper[80], I. Katkov[35], B. Keilhauer[36], E. Kemp[17], R.M. Kieckhafer[82], H.O. Klages[36], M. Kleifges[37], J. Kleinfeller[9], R. Krause[39], N. Krohm[34], D. Kuempel[39], G. Kukec Mezek[71], N. Kunka[37], A.W. Kuotb Awad[36], D. LaHurd[76], L. Latronico[52], R. Lauer[90], M. Lauscher[39], P. Lautridou[33], S. Le Coz[31], D. Lebrun[31], P. Lebrun[80], M.A. Leigui de Oliveira[21], A. Letessier-Selvon[30], I. Lhenry-Yvon[28], K. Link[35], L. Lopes[65], R. López[53], A. López Casado[75], K. Louedec[31], A. Lucero[8], M. Malacari[12], M. Mallamaci[42], J. Maller[33], D. Mandat[26], P. Mantsch[80], A.G. Mariazzi[4], V. Marin[33], I.C. Mariş[74], G. Marsella[47], D. Martello[47], H. Martinez[54], O. Martínez Bravo[53], D. Martraire[28], J.J. Masías Meza[3], H.J. Mathes[36], S. Mathys[34], J. Matthews[81], J.A.J. Matthews[90], G. Matthiae[44], D. Maurizio[13], E. Mayotte[77], P.O. Mazur[80], C. Medina[77], G. Medina-Tanco[58], R. Meissner[39], V.B.B. Mello[22], D. Melo[8], A. Menshikov[37], S. Messina[60], M.I. Micheletti[6], L. Middendorf[39], I.A. Minaya[72], L. Miramonti[42], B. Mitrica[66], L. Molina-Bueno[74], S. Mollerach[1], F. Montanet[31], C. Morello[52], M. Mostafá[86], C.A. Moura[21], G. Müller[39], M.A. Muller[17,20], S. Müller[36], S. Navas[74], P. Necesal[26], L. Nellen[58], A. Nelles[59,61], J. Neuser[34], P.H. Nguyen[12], M. Niculescu-Oglinzanu[66], M. Niechciol[41], L. Niemietz[34], T. Niggemann[39], D. Nitz[82], D. Nosek[25], V. Novotny[25], L. Nožka[27], L.A. Núñez[24], L. Ochilo[41], F. Oikonomou[86], A. Olinto[87], N. Pacheco[73], D. Pakk Selmi-Dei[17], M. Palatka[26], J. Pallotta[2], P. Papenbreer[34], G. Parente[75], A. Parra[53], T. Paul[79,84], M. Pech[26], J. Pękala[63], R. Pelayo[55], I.M. Pepe[19], L. Perrone[47], E. Petermann[89], C. Peters[39], S. Petrera[48,49], Y. Petrov[78], J. Phuntsok[86], R. Piegaia[3], T. Pierog[36], P. Pieroni[3], M. Pimenta[65], V. Pirronello[45], M. Platino[8], M. Plum[39], A. Porcelli[36], C. Porowski[63], R.R. Prado[15], P. Privitera[87], M. Prouza[26], E.J. Quel[2], S. Querchfeld[34], S. Quinn[76], J. Rautenberg[34], O. Ravel[33], D. Ravignani[8], D. Reinert[39], B. Revenu[33], J. Ridky[26], M. Risse[41], P. Ristori[2], V. Rizi[48], W. Rodrigues de Carvalho[75], J. Rodriguez

Rojo[9], M.D. Rodríguez-Frías[73], D. Rogozin[36], J. Rosado[72], M. Roth[36], E. Roulet[1], A.C. Rovero[5], S.J. Saffi[12], A. Saftoiu[66], F. Salamida[28,c], H. Salazar[53], A. Saleh[71], F. Salesa Greus[86], G. Salina[44], J.D. Sanabria Gomez[24], F. Sánchez[8], P. Sanchez-Lucas[74], E.M. Santos[16], E. Santos[17], F. Sarazin[77], B. Sarkar[34], R. Sarmento[65], C. Sarmiento-Cano[24], R. Sato[9], C. Scarso[9], M. Schauer[34], V. Scherini[47], H. Schieler[36], D. Schmidt[36], O. Scholten[60,b], H. Schoorlemmer[88], P. Schovánek[26], F.G. Schröder[36], A. Schulz[36], J. Schulz[59], J. Schumacher[39], S.J. Sciutto[4], A. Segreto[50], M. Settimo[30], A. Shadkam[81], R.C. Shellard[13], G. Sigl[40], O. Sima[68], A. Śmiałkowski[64], R. Šmída[36], G.R. Snow[89], P. Sommers[86], S. Sonntag[41], J. Sorokin[12], R. Squartini[9], Y.N. Srivastava[84], D. Stanca[66], S. Stanič[71], J. Stapleton[85], J. Stasielak[63], M. Stephan[39], A. Stutz[31], F. Suarez[8,11], M. Suarez Durán[24], T. Suomijärvi[28], A.D. Supanitsky[5], M.S. Sutherland[85], J. Swain[84], Z. Szadkowski[64], O.A. Taborda[1], A. Tapia[8], A. Tepe[41], V.M. Theodoro[17], O. Tibolla[56], C. Timmermans[59,61], C.J. Todero Peixoto[14], G. Toma[66], L. Tomankova[36], B. Tomé[65], A. Tonachini[46], G. Torralba Elipe[75], D. Torres Machado[22], P. Travnicek[26], M. Trini[71], R. Ulrich[36], M. Unger[83,36], M. Urban[39], J.F. Valdés Galicia[58], I. Valiño[75], L. Valore[43], G. van Aar[59], P. van Bodegom[12], A.M. van den Berg[60], S. van Velzen[59], A. van Vliet[40], E. Varela[53], B. Vargas Cárdenas[58], G. Varner[88], R. Vasquez[22], J.R. Vázquez[72], R.A. Vázquez[75], D. Veberič[36], V. Verzi[44], J. Vicha[26], M. Videla[8], L. Villaseñor[57], B. Vlcek[73], S. Vorobiov[71], H. Wahlberg[4], O. Wainberg[8,11], D. Walz[39], A.A. Watson[a], M. Weber[37], K. Weidenhaupt[39], A. Weindl[36], C. Welling[39], F. Werner[35], A. Widom[84], L. Wiencke[77], H. Wilczyński[63], T. Winchen[34], D. Wittkowski[34], B. Wundheiler[8], S. Wykes[59], L. Yang[71], T. Yapici[82], A. Yushkov[41], E. Zas[75], D. Zavrtanik[71,70], M. Zavrtanik[70,71], A. Zepeda[54], B. Zimmermann[37], M. Ziolkowski[41], F. Zuccarello[45]

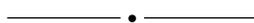

[1] Centro Atómico Bariloche and Instituto Balseiro (CNEA-UNCuyo-CONICET), San Carlos de Bariloche, Argentina
[2] Centro de Investigaciones en Láseres y Aplicaciones, CITEDEF and CONICET, Villa Martelli, Argentina
[3] Departamento de Física, FCEyN, Universidad de Buenos Aires and CONICET, Buenos Aires, Argentina
[4] IFLP, Universidad Nacional de La Plata and CONICET, La Plata, Argentina
[5] Instituto de Astronomía y Física del Espacio (IAFE, CONICET-UBA), Buenos Aires, Argentina
[6] Instituto de Física de Rosario (IFIR) – CONICET/U.N.R. and Facultad de Ciencias Bioquímicas y Farmacéuticas U.N.R., Rosario, Argentina
[7] Instituto de Tecnologías en Detección y Astropartículas (CNEA, CONICET, UNSAM), and Universidad Tecnológica Nacional – Facultad Regional Mendoza (CONICET/CNEA), Mendoza, Argentina
[8] Instituto de Tecnologías en Detección y Astropartículas (CNEA, CONICET, UNSAM), Buenos Aires, Argentina
[9] Observatorio Pierre Auger, Malargüe, Argentina
[10] Observatorio Pierre Auger and Comisión Nacional de Energía Atómica, Malargüe, Argentina
[11] Universidad Tecnológica Nacional – Facultad Regional Buenos Aires, Buenos Aires, Argentina
[12] University of Adelaide, Adelaide, S.A., Australia
[13] Centro Brasileiro de Pesquisas Fisicas, Rio de Janeiro, RJ, Brazil
[14] Universidade de São Paulo, Escola de Engenharia de Lorena, Lorena, SP, Brazil
[15] Universidade de São Paulo, Instituto de Física de São Carlos, São Carlos, SP, Brazil
[16] Universidade de São Paulo, Instituto de Física, São Paulo, SP, Brazil
[17] Universidade Estadual de Campinas, IFGW, Campinas, SP, Brazil
[18] Universidade Estadual de Feira de Santana, Feira de Santana, Brazil
[19] Universidade Federal da Bahia, Salvador, BA, Brazil
[20] Universidade Federal de Pelotas, Pelotas, RS, Brazil
[21] Universidade Federal do ABC, Santo André, SP, Brazil
[22] Universidade Federal do Rio de Janeiro, Instituto de Física, Rio de Janeiro, RJ, Brazil

[23] Universidade Federal Fluminense, EEIMVR, Volta Redonda, RJ, Brazil
[24] Universidad Industrial de Santander, Bucaramanga, Colombia
[25] Charles University, Faculty of Mathematics and Physics, Institute of Particle and Nuclear Physics, Prague, Czech Republic
[26] Institute of Physics of the Academy of Sciences of the Czech Republic, Prague, Czech Republic
[27] Palacky University, RCPTM, Olomouc, Czech Republic
[28] Institut de Physique Nucléaire d'Orsay (IPNO), Université Paris 11, CNRS-IN2P3, Orsay, France
[29] Laboratoire de l'Accélérateur Linéaire (LAL), Université Paris 11, CNRS-IN2P3, Orsay, France
[30] Laboratoire de Physique Nucléaire et de Hautes Energies (LPNHE), Universités Paris 6 et Paris 7, CNRS-IN2P3, Paris, France
[31] Laboratoire de Physique Subatomique et de Cosmologie (LPSC), Université Grenoble-Alpes, CNRS/IN2P3, Grenoble, France
[32] Station de Radioastronomie de Nançay, Observatoire de Paris, CNRS/INSU, Nançay, France
[33] SUBATECH, École des Mines de Nantes, CNRS-IN2P3, Université de Nantes, Nantes, France
[34] Bergische Universität Wuppertal, Fachbereich C – Physik, Wuppertal, Germany
[35] Karlsruhe Institute of Technology – Campus South – Institut für Experimentelle Kernphysik (IEKP), Karlsruhe, Germany
[36] Karlsruhe Institute of Technology – Campus North – Institut für Kernphysik, Karlsruhe, Germany
[37] Karlsruhe Institute of Technology – Campus North – Institut für Prozessdatenverarbeitung und Elektronik, Karlsruhe, Germany
[38] Max-Planck-Institut für Radioastronomie, Bonn, Germany
[39] RWTH Aachen University, III. Physikalisches Institut A, Aachen, Germany
[40] Universität Hamburg, II. Institut für Theoretische Physik, Hamburg, Germany
[41] Universität Siegen, Fachbereich 7 Physik – Experimentelle Teilchenphysik, Siegen, Germany
[42] Università di Milano and Sezione INFN, Milan, Italy
[43] Università di Napoli "Federico II" and Sezione INFN, Napoli, Italy
[44] Università di Roma II "Tor Vergata" and Sezione INFN, Roma, Italy
[45] Università di Catania and Sezione INFN, Catania, Italy
[46] Università di Torino and Sezione INFN, Torino, Italy
[47] Dipartimento di Matematica e Fisica "E. De Giorgi" dell'Università del Salento and Sezione INFN, Lecce, Italy
[48] Dipartimento di Scienze Fisiche e Chimiche dell'Università dell'Aquila and Sezione INFN, L'Aquila, Italy
[49] Gran Sasso Science Institute (INFN), L'Aquila, Italy
[50] Istituto di Astrofisica Spaziale e Fisica Cosmica di Palermo (INAF), Palermo, Italy
[51] INFN, Laboratori Nazionali del Gran Sasso, Assergi (L'Aquila), Italy
[52] Osservatorio Astrofisico di Torino (INAF), Università di Torino and Sezione INFN, Torino, Italy
[53] Benemérita Universidad Autónoma de Puebla, Puebla, México
[54] Centro de Investigación y de Estudios Avanzados del IPN (CINVESTAV), México, D.F., México
[55] Unidad Profesional Interdisciplinaria en Ingeniería y Tecnologías Avanzadas del Instituto Politécnico Nacional (UPIITA-IPN), México, D.F., México
[56] Universidad Autónoma de Chiapas, Tuxtla Gutiérrez, Chiapas, México
[57] Universidad Michoacana de San Nicolás de Hidalgo, Morelia, Michoacán, México
[58] Universidad Nacional Autónoma de México, México, D.F., México
[59] IMAPP, Radboud University Nijmegen, Nijmegen, Netherlands
[60] KVI – Center for Advanced Radiation Technology, University of Groningen, Groningen, Netherlands
[61] Nikhef, Science Park, Amsterdam, Netherlands
[62] ASTRON, Dwingeloo, Netherlands
[63] Institute of Nuclear Physics PAN, Krakow, Poland
[64] University of Łódź, Łódź, Poland
[65] Laboratório de Instrumentação e Física Experimental de Partículas (LIP) and Instituto Superior Técnico, Universidade de Lisboa (UL), Portugal

[66] "Horia Hulubei" National Institute for Physics and Nuclear Engineering, Bucharest-Magurele, Romania
[67] Institute of Space Science, Bucharest-Magurele, Romania
[68] University of Bucharest, Physics Department, Bucharest, Romania
[69] University Politehnica of Bucharest, Bucharest, Romania
[70] Experimental Particle Physics Department, J. Stefan Institute, Ljubljana, Slovenia
[71] Laboratory for Astroparticle Physics, University of Nova Gorica, Nova Gorica, Slovenia
[72] Universidad Complutense de Madrid, Madrid, Spain
[73] Universidad de Alcalá, Alcalá de Henares, Madrid, Spain
[74] Universidad de Granada and C.A.F.P.E., Granada, Spain
[75] Universidad de Santiago de Compostela, Santiago de Compostela, Spain
[76] Case Western Reserve University, Cleveland, OH, USA
[77] Colorado School of Mines, Golden, CO, USA
[78] Colorado State University, Fort Collins, CO, USA
[79] Department of Physics and Astronomy, Lehman College, City University of New York, Bronx, NY, USA
[80] Fermilab, Batavia, IL, USA
[81] Louisiana State University, Baton Rouge, LA, USA
[82] Michigan Technological University, Houghton, MI, USA
[83] New York University, New York, NY, USA
[84] Northeastern University, Boston, MA, USA
[85] Ohio State University, Columbus, OH, USA
[86] Pennsylvania State University, University Park, PA, USA
[87] University of Chicago, Enrico Fermi Institute, Chicago, IL, USA
[88] University of Hawaii, Honolulu, HI, USA
[89] University of Nebraska, Lincoln, NE, USA
[90] University of New Mexico, Albuquerque, NM, USA
[a] School of Physics and Astronomy, University of Leeds, Leeds, United Kingdom
[b] Also at Vrije Universiteit Brussels, Brussels, Belgium
[c] Currently at INFN Milano Bicocca, Milano, Italy

# The Telescope Array Collaboration

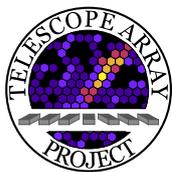

R.U. Abbasi[1], M. Abe[2], T. Abu-Zayyad[1], M. Allen[1], R. Azuma[3], E. Barcikowski[1], J.W. Belz[1], D.R. Bergman[1], S.A. Blake[1], R. Cady[1], M.J. Chae[4], B.G. Cheon[5], J. Chiba[6], M. Chikawa[7], W.R. Cho[8], T. Fujii[9], M. Fukushima[9,10], T. Goto[11], W. Hanlon[1], Y. Hayashi[11], N. Hayashida[12], K. Hibino[12], K. Honda[13], D. Ikeda[9], N. Inoue[2], T. Ishii[13], R. Ishimori[3], H. Ito[14], D. Ivanov[1], C.C.H. Jui[1], K. Kadota[15], F. Kakimoto[3], O. Kalashev[16], K. Kasahara[17], H. Kawai[18], S. Kawakami[11], S. Kawana[2], K. Kawata[9], E. Kido[9], H.B. Kim[5], J.H. Kim[1], J.H. Kim[19], S. Kitamura[3], Y. Kitamura[3], V. Kuzmin[16†], Y.J. Kwon[8], J. Lan[1], S.I. Lim[4], J.P. Lundquist[1], K. Machida[13], K. Martens[10], T. Matsuda[20], T. Matsuyama[11], J.N. Matthews[1], M. Minamino[11], Y. Mukai[13], I. Myers[1], K. Nagasawa[2], S. Nagataki[14], T. Nakamura[21], T. Nonaka[9], A. Nozato[7], S. Ogio[11], J. Ogura[3], M. Ohnishi[9], H. Ohoka[9], K. Oki[9], T. Okuda[22], M. Ono[23], A. Oshima[24], S. Ozawa[17], I.H. Park[25], M.S. Pshirkov[16,26], D.C. Rodriguez[1], G. Rubtsov[16], D. Ryu[19], H. Sagawa[9], N. Sakurai[11], L.M. Scott[27], P.D. Shah[1], F. Shibata[13], T. Shibata[9], H. Shimodaira[9], B.K. Shin[5], H.S. Shin[9], J.D. Smith[1], P. Sokolsky[1], R.W. Springer[1], B.T. Stokes[1], S.R. Stratton[1,27], T.A. Stroman[1], T. Suzawa[2], M. Takamura[6], M. Takeda[9], R. Takeishi[9], A. Taketa[28], M. Takita[9], Y. Tameda[12], H. Tanaka[11], K. Tanaka[29], M. Tanaka[20], S.B. Thomas[1], G.B. Thomson[1], P. Tinyakov[30,16], I. Tkachev[16], H. Tokuno[3], T. Tomida[31], S. Troitsky[16], Y. Tsunesada[3], K. Tsutsumi[3], Y. Uchihori[32], S. Udo[12], F. Urban[30], G. Vasiloff[1], T. Wong[1], R. Yamane[11], H. Yamaoka[20], K. Yamazaki[28], J. Yang[4], K. Yashiro[6], Y. Yoneda[11], S. Yoshida[18], H. Yoshii[33], R. Zollinger[1], Z. Zundel[1]

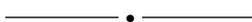

[1] High Energy Astrophysics Institute and Department of Physics and Astronomy, University of Utah, Salt Lake City, Utah, USA
[2] The Graduate School of Science and Engineering, Saitama University, Saitama, Saitama, Japan
[3] Graduate School of Science and Engineering, Tokyo Institute of Technology, Meguro, Tokyo, Japan
[4] Department of Physics and Institute for the Early Universe, Ewha Womans University, Seodaaemun-gu, Seoul, Korea
[5] Department of Physics and The Research Institute of Natural Science, Hanyang University, Seongdong-gu, Seoul, Korea
[6] Department of Physics, Tokyo University of Science, Noda, Chiba, Japan
[7] Department of Physics, Kinki University, Higashi Osaka, Osaka, Japan
[8] Department of Physics, Yonsei University, Seodaemun-gu, Seoul, Korea
[9] Institute for Cosmic Ray Research, University of Tokyo, Kashiwa, Chiba, Japan
[10] Kavli Institute for the Physics and Mathematics of the Universe (WPI), Todai Institutes for Advanced Study, the University of Tokyo, Kashiwa, Chiba, Japan
[11] Graduate School of Science, Osaka City University, Osaka, Osaka, Japan
[12] Faculty of Engineering, Kanagawa University, Yokohama, Kanagawa, Japan
[13] Interdisciplinary Graduate School of Medicine and Engineering, University of Yamanashi, Kofu, Yamanashi, Japan
[14] Astrophysical Big Bang Laboratory, RIKEN, Wako, Saitama, Japan
[15] Department of Physics, Tokyo City University, Setagaya-ku, Tokyo, Japan
[16] Institute for Nuclear Research of the Russian Academy of Sciences, Moscow, Russia
[17] Advanced Research Institute for Science and Engineering, Waseda University, Shinjuku-ku, Tokyo, Japan
[18] Department of Physics, Chiba University, Chiba, Chiba, Japan
[19] Department of Physics, School of Natural Sciences, Ulsan National Institute of Science and Technology, UNIST-gil, Ulsan, Korea
[20] Institute of Particle and Nuclear Studies, KEK, Tsukuba, Ibaraki, Japan
[21] Faculty of Science, Kochi University, Kochi, Kochi, Japan

[22] Department of Physical Sciences, Ritsumeikan University, Kusatsu, Shiga, Japan
[23] Department of Physics, Kyushu University, Fukuoka, Fukuoka, Japan
[24] Engineering Science Laboratory, Chubu University, Kasugai, Aichi, Japan
[25] Department of Physics, Sungkyunkwan University, Jang-an-gu, Suwon, Korea
[26] Sternberg Astronomical Institute, Moscow M.V. Lomonosov State University, Moscow, Russia
[27] Department of Physics and Astronomy, Rutgers University – The State University of New Jersey, Piscataway, New Jersey, USA
[28] Earthquake Research Institute, University of Tokyo, Bunkyo-ku, Tokyo, Japan
[29] Graduate School of Information Sciences, Hiroshima City University, Hiroshima, Hiroshima, Japan
[30] Service de Physique Théorique, Université Libre de Bruxelles, Brussels, Belgium
[31] Department of Computer Science and Engineering, Shinshu University, Nagano, Nagano, Japan
[32] National Institute of Radiological Science, Chiba, Chiba, Japan
[33] Department of Physics, Ehime University, Matsuyama, Ehime, Japan
[†] Deceased.

# Acknowledgments of the IceCube Collaboration

We acknowledge the support from the following agencies: U.S. National Science Foundation-Office of Polar Programs, U.S. National Science Foundation-Physics Division, University of Wisconsin Alumni Research Foundation, the Grid Laboratory Of Wisconsin (GLOW) grid infrastructure at the University of Wisconsin - Madison, the Open Science Grid (OSG) grid infrastructure; U.S. Department of Energy, and National Energy Research Scientific Computing Center, the Louisiana Optical Network Initiative (LONI) grid computing resources; Natural Sciences and Engineering Research Council of Canada, WestGrid and Compute/Calcul Canada; Swedish Research Council, Swedish Polar Research Secretariat, Swedish National Infrastructure for Computing (SNIC), and Knut and Alice Wallenberg Foundation, Sweden; German Ministry for Education and Research (BMBF), Deutsche Forschungsgemeinschaft (DFG), Helmholtz Alliance for Astroparticle Physics (HAP), Research Department of Plasmas with Complex Interactions (Bochum), Germany; Fund for Scientific Research (FNRS-FWO), FWO Odysseus programme, Flanders Institute to encourage scientific and technological research in industry (IWT), Belgian Federal Science Policy Office (Belspo); University of Oxford, United Kingdom; Marsden Fund, New Zealand; Australian Research Council; Japan Society for Promotion of Science (JSPS); the Swiss National Science Foundation (SNSF), Switzerland; National Research Foundation of Korea (NRF); Danish National Research Foundation, Denmark (DNRF).

# Acknowledgments of the Pierre Auger Collaboration

The successful installation, commissioning, and operation of the Pierre Auger Observatory would not have been possible without the strong commitment and effort from the technical and administrative staff in Malargüe. We are very grateful to the following agencies and organizations for financial support:

Comisión Nacional de Energía Atómica, Agencia Nacional de Promoción Científica y Tecnológica (ANPCyT), Consejo Nacional de Investigaciones Científicas y Técnicas (CONICET), Gobierno de la Provincia de Mendoza, Municipalidad de Malargüe, NDM Holdings and Valle Las Leñas, in gratitude for their continuing cooperation over land access, Argentina; the Australian Research Council; Conselho Nacional de Desenvolvimento Científico e Tecnológico (CNPq), Financiadora de Estudos e Projetos (FINEP), Fundação de Amparo à Pesquisa do Estado de Rio de Janeiro (FAPERJ), São Paulo Research Foundation (FAPESP) Grants No. 2010/07359-6 and No. 1999/05404-3, Ministério de Ciência e Tecnologia (MCT), Brazil; Grant No. MSMT-CR LG13007, No. 7AMB14AR005, and the Czech Science Foundation Grant No. 14-17501S, Czech Republic; Centre de Calcul IN2P3/CNRS, Centre National de la Recherche Scientifique (CNRS), Conseil Régional Ile-de-France, Département Physique Nucléaire et Corpusculaire (PNC-IN2P3/CNRS), Département Sciences de l'Univers (SDU-INSU/CNRS), Institut Lagrange de Paris (ILP) Grant No. LABEX ANR-10-LABX-63, within the Investissements d'Avenir Programme Grant No. ANR-11-IDEX-0004-02, France; Bundesministerium für Bildung und Forschung (BMBF), Deutsche Forschungsgemeinschaft (DFG), Finanzministerium Baden-Württemberg, Helmholtz Alliance for Astroparticle Physics (HAP), Helmholtz-Gemeinschaft Deutscher Forschungszentren (HGF), Ministerium für Wissenschaft und Forschung, Nordrhein Westfalen, Ministerium für Wissenschaft, Forschung und Kunst, Baden-Württemberg, Germany; Istituto Nazionale di Fisica Nucleare (INFN), Istituto Nazionale di Astrofisica (INAF), Ministero dell'Istruzione, dell'Universitá e della Ricerca (MIUR), Gran Sasso Center for Astroparticle Physics (CFA), CETEMPS Center of Excellence, Ministero degli Affari Esteri (MAE), Italy; Consejo Nacional de Ciencia y Tecnología (CONACYT), Mexico; Ministerie van Onderwijs, Cultuur en Wetenschap, Nederlandse Organisatie voor Wetenschappelijk Onderzoek (NWO), Stichting voor Fundamenteel Onderzoek der Materie (FOM), Netherlands; National Centre for Research and Development, Grants No. ERA-NET-ASPERA/01/11 and No. ERA-NET-ASPERA/02/11, National Science Centre, Grants No. 2013/08/M/ST9/00322, No. 2013/08/M/ST9/00728 and No. HARMONIA 5 - 2013/10/M/ST9/00062, Poland; Portuguese national funds and FEDER funds within Programa Operacional Factores de Competitividade through Fundação para a Ciência e a Tecnologia (COMPETE), Portugal; Romanian Authority for Scientific Research ANCS, CNDI-UEFISCDI partnership projects Grants No. 20/2012 and No. 194/2012, Grants No. 1/AS-PERA2/2012 ERA-NET, No. PN-II-RU-PD-2011-3-0145-17 and No. PN-II-RU-PD-2011-3-0062, the Minister of National Education, Programme Space Technology and Advanced Research (STAR), Grant No. 83/2013, Romania; Slovenian Research Agency, Slovenia; Comunidad de Madrid, FEDER funds, Ministerio de Educación y Ciencia, Xunta de Galicia, European Community 7th Framework Program, Grant No. FP7-PEOPLE-2012-IEF-328826, Spain; Science and Technology Facilities Council, United Kingdom; Department of Energy, Contracts No. DE-AC02-07CH11359, No. DE-FR02-04ER41300, No. DE-FG02-99ER41107 and No. DE-SC0011689, National Science Foundation, Grant No. 0450696, The Grainger Foundation, USA; NAFOSTED, Vietnam; Marie Curie-IRSES/EPLANET, European Particle Physics Latin American Network, European Union 7th Framework Program, Grant No. PIRSES-2009-GA-246806 and PIOF-GA-2013-624803; and UNESCO.

## Acknowledgments of the Telescope Array Collaboration

The Telescope Array experiment is supported by the Japan Society for the Promotion of Science through Grants-in-Aid for Scientific Research on Specially Promoted Research (21000002) "Extreme Phenomena in the Universe Explored by Highest Energy Cosmic Rays" and for Scientific Research (19104006), and the Inter-University Research Program of the Institute for Cosmic Ray Research; by the U.S. National Science Foundation awards PHY-0307098, PHY-0601915, PHY-0649681, PHY-0703893, PHY-0758342, PHY-0848320, PHY-1069280, PHY-1069286, PHY-1404495 and PHY-1404502; by the National Research Foundation of Korea (2007-0093860, R32-10130, 2012R1A1A2008381, 2013004883); by the Russian Academy of Sciences, RFBR grants 11-02-01528a and 13-02-01311a (INR), IISN project No. 4.4502.13, and Belgian Science Policy under IUAP VII/37 (ULB). The foundations of Dr. Ezekiel R. and Edna Wattis Dumke, Willard L. Eccles, and George S. and Dolores Doré Eccles all helped with generous donations. The State of Utah supported the project through its Economic Development Board, and the University of Utah through the Office of the Vice President for Research. The experimental site became available through the cooperation of the Utah School and Institutional Trust Lands Administration (SITLA), U.S. Bureau of Land Management, and the U.S. Air Force. We also wish to thank the people and the officials of Millard County, Utah for their steadfast and warm support. We gratefully acknowledge the contributions from the technical staffs of our home institutions. An allocation of computer time from the Center for High Performance Computing at the University of Utah is gratefully acknowledged.

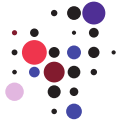

# Search for a correlation between the UHECRs measured by the Pierre Auger Observatory and the Telescope Array and the neutrino candidate events from IceCube

**The IceCube Collaboration**[1]**, The Pierre Auger Collaboration**[2]**, The Telescope Array Collaboration**[3]

[1] http://icecube.wisc.edu/collaboration/authors/icrc15_icecube
[2] http://www.auger.org/archive/authors_2015_ICRC.html
[3] http://www.telescopearray.org/index.php/research/collaborators
*E-mail:* christov@cern.ch

We have conducted three searches for correlations between ultra-high energy cosmic rays detected by the Telescope Array and the Pierre Auger Observatory, and high-energy neutrino candidate events from IceCube. Two cross-correlation analyses with UHECRs are done: one with 39 cascades from the IceCube 'high-energy starting events' sample and the other with 16 high-energy 'track events'. The angular separation between the arrival directions of neutrinos and UHECRs is scanned over. The same events are also used in a separate search using a maximum likelihood approach, after the neutrino arrival directions are stacked. To estimate the significance we assume UHECR magnetic deflections to be inversely proportional to their energy, with values $3°$, $6°$ and $9°$ at $100\,\text{EeV}$ to allow for the uncertainties on the magnetic field strength and UHECR charge. A similar analysis is performed on stacked UHECR arrival directions and the IceCube sample of through-going muon track events which were optimized for neutrino point-source searches.

**Corresponding authors:** A. Christov[4], G. Golup[*,5], J. Aublin[6], L. Caccianiga[6], P.L. Ghia[6], T. Montaruli[4], M. Rameez[4], E. Roulet[5], H. Sagawa[7], P. Tinyakov[8], and M. Unger[9]

[4] Départment de physique nucléaire et corpusculaire, Université de Genève, 24 Quai Ernest Ansermet, 1211 Genève, Switzerland.
[5] Centro Atómico Bariloche, Av. Bustillo 9500, S. C. de Bariloche 8400, Argentina.
[6] Laboratoire de Physique Nucléaire et de Hautes Energies (LPNHE), Universités Paris 6 et Paris 7, CNRS-IN2P3, 4 place Jussieu, 75252, Paris, France.
[7] Institute for Cosmic Ray Research, University of Tokyo, Kashiwa, Chiba, Japan.
[8] Service de Physique Théorique, Université Libre de Bruxelles, Boulevard du Triomphe (Campus de la Plaine), Ixelles 1050, Belgium.
[9] Karlsruhe Institute of Technology - Campus North - Institut für Kernphysik, Karlsruhe, Germany and New York University, New York, USA.

---

*Speaker.

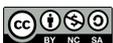

## 1. Introduction

A multi-messenger approach can help to identify the sources of ultra-high energy cosmic rays (UHECRs). It is difficult to do so from their arrival directions since CRs are charged particles so are deflected en-route to Earth. This deflection cannot be computed precisely since the CR composition at ultra-high energies as well as the intervening magnetic field strength are poorly known. If the CR composition is light i.e. mainly protons, the magnetic deflection may be only a few degrees above a few tens of EeV. Secondary particles including neutrinos (νs) are produced in the sources by the interactions between the CRs and ambient photon and matter fields. Neutrinos have no charge and interact only through the weak force, so their arrival directions do point back to where they originated from, although they are also hard to detect for the same reason. In this work we describe a joint analysis by the IceCube, Pierre Auger and Telescope Array Collaborations to search for angular correlations between the arrival directions of high-energy νs and UHECRs that would provide insight into the long-standing open question of cosmic ray origin.

## 2. The observatories and data sets

### 2.1 The IceCube Neutrino Telescope

IceCube is a cubic-kilometer neutrino detector installed in the ice at the geographic South Pole [1] between depths of 1450 m and 2450 m. Neutrino reconstruction relies on the optical detection of Cherenkov radiation emitted by secondary particles produced in ν interactions in the surrounding ice or the nearby bedrock.

Depending on the flavor of the interacting neutrino and the type of interaction, different signatures are expected in the detector. The one with the best angular resolution ($\sim 1°$) is the charged current $\nu_\mu$ interaction where a track is produced as the outgoing muon traverses the detector. Cascades are produced in the detector as a result of charged current $\nu_{e,\tau}$ interactions or all neutral current neutrino interactions. In this case the angular resolution is poorer (around 15° above 100 TeV). The resolution of the deposited energy for tracks and cascades is around 15% [2] but cascades have a better resolution for the reconstructed neutrino energy since most of the energy is deposited in the detector, which is not the case for tracks.

Different data sets are considered in this work. A set of cascades that have been detected in a search for high-energy events where the interaction occurs within the detector is used [3]. This set of 39 cascades, which is part of the HESE ('High-Energy Starting Events') set, consists of data taken between May 2010 and May 2014 and is called 'high-energy cascades' in the following (deposited energy range: $\sim 30 - 2000$ TeV). A second set of events referred to as 'high-energy tracks' (energy above $\sim 70$ TeV) is formed by two parts. The first part is the 7 tracks in the HESE sample [3] that have energies and directions which make them more likely to be of extraterrestrial origin than the other track events in that sample. The second part is 9 muon tracks found in a search of a diffuse up-going $\nu_\mu$ flux [4]. These 9 muon tracks, found in two years of data (May 2010-May 2012), belong to a high energy-proxy excess with respect to atmospheric predictions. This excess is compatible with an astrophysical $E^{-2}$ flux at the level indicated by the HESE analysis [4].

The third data set used is called '4-year point-source sample' [5] and consists of events with sub-degree median angular resolution detected between May 2008 and May 2012. The set includes about 400,000 events, mostly up-going atmospheric νs from the Northern hemisphere and high-energy atmospheric muons from the Southern hemisphere.

### 2.2 The Pierre Auger Observatory

The Pierre Auger Observatory is located in Malargüe, Argentina (35.2° S, 69.5° W, 1400 m a.s.l.) [6]. It consists of a surface array of 1660 water-Cherenkov detectors covering an area of approximately 3000 km$^2$. The array is overlooked by 27 telescopes at four sites which constitute the fluorescence detector. The surface and air fluorescence detectors are designed to perform complementary measurements of air showers created by UHECRs.

The data set used for the present analysis includes 231 events with $E > 52$ EeV and zenith angles smaller than 80° recorded by the surface detector array from January 2004 to March 2014 [7]. The exposure determined by geometrical considerations for the period analyzed amounts to 66,452 km$^2$ sr yr. The angular resolution, defined as the 68th percentile of the distribution of opening angles between the true and reconstructed directions of simulated events, is better than 0.9° [8]. The absolute energy scale, given by the fluorescence calibration, has a systematic uncertainty of 14% and the energy resolution is about 12% [9].

### 2.3 Telescope Array

The Telescope Array (TA) is located in Utah, USA (39.3°N, 112.9°W, 1400 m a.s.l.) [10] and detects extensive air showers generated by UHECRs. It comprises a 700 km$^2$ surface array of 507 plastic scintillation detectors, 3 m$^2$ each, distributed in a square grid with 1.2 km spacing. The array is overlooked by 3 fluorescence detector stations with 38 telescopes.

The UHECR sample considered in the present analysis consists of 87 events with $E > 57$ EeV and zenith angles smaller than 55° collected between May 2008 and May 2014 by the surface detector. A subset of events has been published in [11]. The total exposure is around 9,500 km$^2$ sr yr. The angular resolution is better than 1.5°. The energy scale of the surface detector is also calibrated with the fluorescence detector. The energy resolution is better than 20% with a systematic uncertainty on the absolute energy scale of 21% [12].

## 3. Data analyses

There are three different analyses which are presented in detail in this Section. A cross-correlation and a stacking likelihood analysis are done on the sample of high-energy cascades and high-energy tracks and the UHECRs detected by Auger and TA. Cascade and track-like events are considered separately since, due to their different angular resolutions, the angular distance at which a signal (if any) can be detected would be different. A third analysis is performed on stacked UHECRs and the IceCube 4-year point-source sample.

The magnetic deflections of CRs have to be accounted for in the likelihood tests. For simplicity, we model individual deflections as a random variable 2-dimensional Gaussian distribution with the energy-dependent standard deviation $\sigma_{\text{MD}}(E_{\text{CR}}) = D \times 100\,\text{EeV}/E_{\text{CR}}$, and we consider the representative values $D = 3°$, $6°$ and $9°$ (the latter is just used for the likelihood test with the high-energy cascades and high-energy tracks). These values are reasonable test values as shown by a backtracking simulation of the detected UHECRs in the galactic magnetic field models of

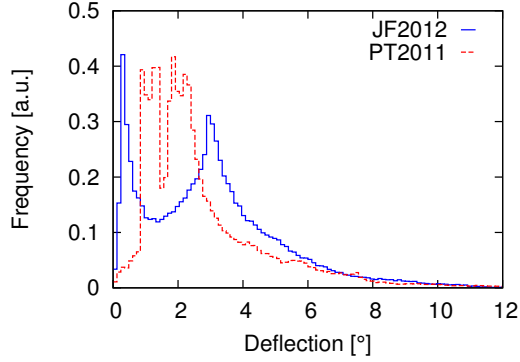

**Figure 1:** Distribution of UHECR deflections in two models for the regular component of the galactic magnetic field, PT2011 [13] and JF2012 [14], for a rigidity $E/Z = 100\,\text{EeV}$.

Pshirkov et al. [13] and Jansson and Farrar [14] and assuming these are protons with $E = 100\,\text{EeV}$. The distributions of the obtained deflections are different for each model (Fig. 1), but the median values for both are $2.7°$. We have then chosen an average value of $3°$. The values of $6°$ and $9°$ are also considered to account for larger deflections that could arise from other light CR components ($Z = 2,3$) or a stronger than predicted strength of the intervening magnetic fields.

### 3.1 UHECR correlation analyses with high-energy cascades and high-energy tracks

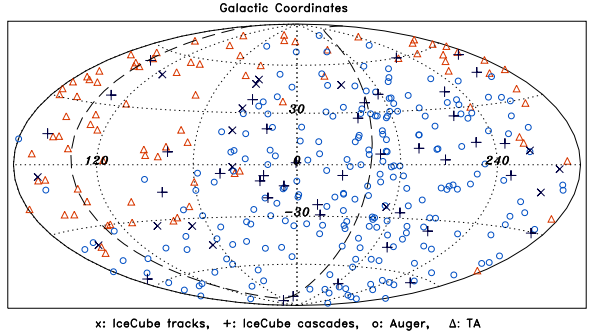

**Figure 2:** Aitoff-Hammer projection of the sky in galactic coordinates showing the arrival directions of the IceCube high-energy cascades (plus signs) and high-energy tracks (crosses), and the UHECRs detected by Auger (circles) and TA (triangles). The dashed line indicates the Super-galactic plane.

The arrival directions of the high-energy tracks and high-energy cascades in IceCube, and of the UHECRs measured by Auger and TA are shown in Fig. 2 in galactic coordinates. Two different analyses are performed with this data set: a cross-correlation and a stacking likelihood analysis.

The cross-correlation method consists of computing the number of UHECR-$\nu$ pairs as a function of their angular separation $\alpha$, $n_p(\alpha)$, and comparing it to the expectation from an isotropic distribution of arrival directions of CRs. The angular scan performed in this case is between $1°$ and $30°$ with a step of $1°$ and, due to this scan, the method does not rely on any assumption about the exact value of the strength of the magnetic deflections, unlike the likelihood method.

In Fig. 3 we show the results obtained applying the cross-correlation method to the data. For the case of the sample of high-energy tracks, the maximum departure from the isotropic expectation of CRs (fixing the positions of the νs) obtained is at an angular distance of 2°, where 1.5 pairs were expected on average and 4 pairs were detected. The post-trial *p*-value is 34%. For the analysis done using the high-energy cascade events, the smallest pre-trial *p*-value occurs at an angular distance of 22°, for which 575 pairs are observed while 490.3 were expected on average. The post-trial *p*-value is $5 \times 10^{-4}$ with respect to expectations of an isotropic flux of CRs. As an a posteriori study, we also evaluated the significance under the hypothesis of an isotropic distribution of neutrinos, fixing the UHECR arrival directions (note that this alternative hypothesis preserves the degree of anisotropy in the arrival directions of CRs that is suggested by the TA 'hot spot' [11] or the excess around Cen A reported by Auger [7]). The post-trial *p*-value is $8.5 \times 10^{-3}$. Thus the cross-correlation of UHECRs with the high-energy cascades provides a potentially interesting result, which we will continue to monitor in the future.

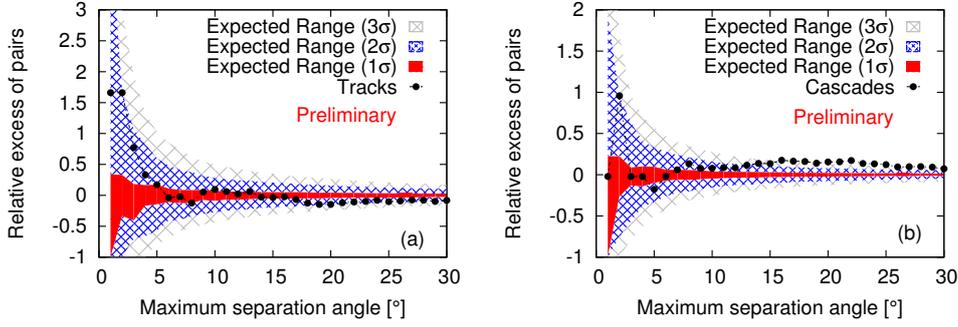

**Figure 3:** Relative excess of pairs, $[n_{\rm p}(\alpha)/\langle n_{\rm p}^{\rm iso}(\alpha)\rangle] - 1$, as a function of the maximum angular separation between the neutrino and UHECR pairs, for the analysis done with the high-energy tracks (a) and with the high-energy cascades (b). The $1\sigma$, $2\sigma$ and $3\sigma$ fluctuations expected from an isotropic distribution of arrival directions of CRs are shown in red, blue and grey, respectively.

Stacking a set of sources is a well known way of accumulating multiple weaker signals to enhance the discovery potential. Since νs are not deflected on their way to Earth, the stacking over sources is replaced by stacking over the set of ν arrival directions. An unbinned likelihood method is used [15], with the log of the likelihood function defined as:

$$\ln \mathscr{L}(n_{\rm s}) = \sum_{i=1}^{N_{\rm Auger}} \ln\left(\frac{n_{\rm s}}{N_{\rm CR}} S^i_{\rm Auger} + \frac{N_{\rm CR} - n_{\rm s}}{N_{\rm CR}} B^i_{\rm Auger}\right) + \sum_{i=1}^{N_{\rm TA}} \ln\left(\frac{n_{\rm s}}{N_{\rm CR}} S^i_{\rm TA} + \frac{N_{\rm CR} - n_{\rm s}}{N_{\rm CR}} B^i_{\rm TA}\right),$$

where $n_{\rm s}$, the number of signal events, is the only free parameter, $N_{\rm CR} = N_{\rm Auger} + N_{\rm TA}$, $S^i_{\rm Auger}$ and $S^i_{\rm TA}$ are the signal PDFs (Probability Distribution Functions) for Auger and for TA, respectively, and $B^i_{\rm Auger}$ and $B^i_{\rm TA}$ are the corresponding background PDFs. The signal PDFs, in which the different neutrino positions are stacked, take into account the exposure and angular resolution of the CR observatories, the assumed CR magnetic deflections and the likelihood maps for the reconstruction of the νs arrival directions (Fig. 4). The background PDFs are the normalized exposures of the CR observatories. The test statistic *TS* is defined as: $TS = -2\ln\frac{\mathscr{L}(n_{\rm s})}{\mathscr{L}(n_{\rm s}=0)}$ and follows a distribution close to $\chi^2$ with one degree of freedom.

The results for the stacking method are shown in Table 1. The most significant deviation from an isotropic flux of CRs occurs for the magnetic deflection parameter $D = 6°$ with the high-energy cascades. The observed pre-trial $p$-value of $2.7 \times 10^{-4}$ corresponds to $8 \times 10^{-4}$ post-trial, i.e. after accounting for the 3 values of $D$ considered. Therefore, we obtain a potentially interesting result with the cascades as in the case of the cross-correlation analysis, which will be further studied with a larger number of events.

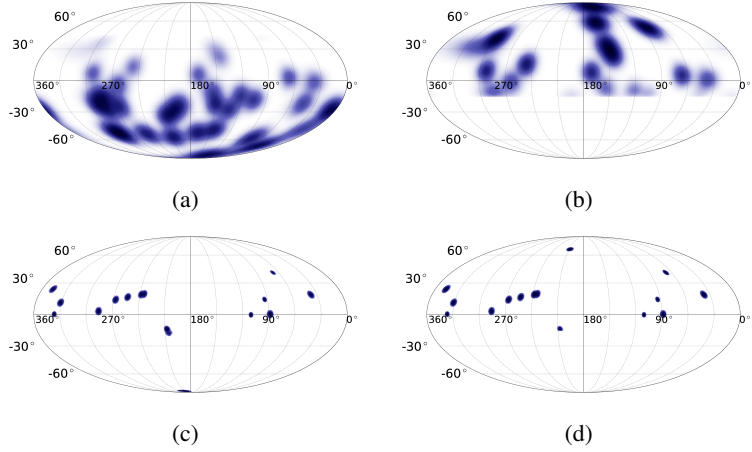

**Figure 4:** The signal PDFs before the Gaussian smearing in equatorial coordinates. The upper plots are for the high-energy cascades, while the lower ones are for the high-energy tracks. The declination-dependent exposure is applied for Auger in the left-hand plots and for TA in the right-hand plots.

| $D$ | High-energy tracks | | High-energy cascades | |
|---|---|---|---|---|
| | $n_s$ | pre-trial $p$-value | $n_s$ | pre-trial $p$-value |
| 3° | 4.3 | 0.22 | 53.7 | $2.1 \times 10^{-3}$ |
| 6° | 0.5 | 0.48 | 85.7 | $2.7 \times 10^{-4}$ |
| 9° | - | underfluctuation | 106.1 | $3.8 \times 10^{-4}$ |

**Table 1:** Results for the stacking analyses with the sample of high-energy tracks and high-energy cascades.

The angular distance at which an excess would occur in the case of the cross-correlation includes not only the magnetic deflections at the corresponding CR energies but also the experimental angular uncertainties. In the case of cascades, the angular uncertainty is $\sim 15°$ and it is $\sim 1°$ for CRs. Since most CRs in the data set have $E_{CR} \sim 60$ EeV, the assumed magnetic deflection where the smallest $p$-value is found in the case of the likelihood analysis with the cascades ($\sigma_{MD}(E_{CR}) = 6 \times 100\,\text{EeV}/E_{CR}$) is $\sim 10°$ in most cases. To translate this into an angular scale where one would find an excess in the cross-correlation analysis (if there were a signal), we add in quadrature and we obtain $\sqrt{(15°)^2 + (1°)^2 + (10°)^2} \sim 18°$. This scale is comparable to the 22° where the smallest $p$-value is found for the cross-correlation performed with the cascades. Hence, the magnetic deflection of the CRs one would infer from the cross-correlation analysis with the cascades is comparable to the one assumed for the smallest $p$-value in the likelihood analysis, even if none of the results are at a level where no strong claims can be made.

### 3.2 Stacking search for neutrino point-sources in the 4 year point-source sample

The νs data set used for this analysis is the IceCube point-source data set. A stacking analysis is done but in this case (as opposed to the previous one) the stacked sources are the measured positions of UHECRs. An unbinned likelihood method is performed where the log likelihood is defined as: $\ln \mathscr{L}(n_{s_\nu}, \gamma) = \sum_{i=1}^{N_\nu} \ln \left( \frac{n_{s_\nu}}{N_\nu} S_i(\gamma, E_i) + \left(1 - \frac{n_{s_\nu}}{N_\nu}\right) B_i \right)$, with $n_{s_\nu}$ the total number of neutrino signal events and $\gamma$ the neutrino spectral index assuming a power-law energy spectrum. The stacked signal PDF is defined as $S_i = \sum_{j=1}^{N_{CR}} R_{IC}(\delta_j, \gamma) S_i^j / \sum_{j=1}^{N_{CR}} R_{IC}(\delta_j, \gamma)$, with $R_{IC}(\delta_j, \gamma)$ the IceCube acceptance at the declination of a CR $j$. The signal PDF is $S_i^j = \frac{1}{2\pi(\sigma_i^2 + \sigma_j^2)} e^{-r_{ij}^2/2(\sigma_i^2 + \sigma_j^2)} P(E_i|\gamma)$, where $r_{ij}$ is the angular distance between the νs and CRs, $\sigma_i$ is the angular resolution of the ν and $P(E_i|\gamma)$ is the energy PDF (function of the reconstructed energy proxy $E_i$ and $\gamma$). The CR deflection is modeled as an extension of the source in the likelihood with $\sigma_j = \sqrt{\sigma_{MD}^2 + \sigma_{exp}^2}$, where $\sigma_{exp} = 0.9°$ or 1.5° is the experimental angular resolution of Auger or TA, respectively. The background PDF is $B_i = B(\theta_i) P_{atm}(E_i)$ where the energy PDF $P_{atm}(E_i)$ represents the probability of obtaining an energy $E_i$ from atmospheric backgrounds. The two free parameters are $n_{s_\nu}$ and $\gamma$.

If we were to consider the entire data set of UHECRs as sources in the likelihood, their total extensions would cover a considerable amount of the sky, reducing the effectiveness of the anisotropy search. Hence we decided to introduce a threshold energy, $E_{th}$, below which the CRs would not be considered. To obtain $E_{th}$, we have performed simulations of ν arrival directions and we have used the real sample of UHECRs, sampling different $E_{th}$ energies. The flux required for a pre-trial $p$-value of $5\sigma$ as a function of $E_{th}$, is shown in Fig. 5. With the objective of keeping the flux required per source for discovery low while keeping as many UHECR events as possible, an energy threshold $E_{th} = 85$ EeV has been adopted. After the application of this cut, 15 CRs in the Northern sky and 12 CRs in the Southern sky remain. Due to the different energy ranges between the neutrino candidate events in the Southern hemisphere ($\sim 100$ TeV – 100 PeV) and in the Northern hemisphere ($\sim 1$ TeV – 1 PeV), for the same number of signal events per source, the normalization of flux required for Northern sources is smaller than for Southern ones. Consequently (and thanks to the gain given from having more stacked sources), the all sky sensitivity is similar to the Northern one (Fig. 5). We have thus not made a distinction between the neutrino sets from each hemisphere in this analysis.

Applying the method to the actual data, all observations are found to be compatible with the background only hypothesis. The smallest post-trial $p$-value is 25% for the hypothesis of $D = 3°$, with a fitted excess of $\sim 123$ events and $\gamma = -3.24$. The analysis with $D = 6°$ yields a $p$-value larger than 50%.

## 4. Conclusions

Three analyses have been performed to investigate correlations between UHECRs detected by the Pierre Auger Observatory and Telescope Array with various samples of IceCube ν candidates. The results we obtained are all below $3.3\sigma$. There is a potentially interesting result in the analyses performed with the set of high-energy cascades when comparing the results to isotropic arrival directions of CRs. If we compare the results to an isotropic flux of neutrinos (fixing the positions of the CRs) to consider the effect of anisotropies in the arrival directions of CRs (such as the TA hot spot), the significance is $\sim 2.4\sigma$. These results were obtained with relatively few events and we will update these analyses in the future with further statistics to follow their evolution.

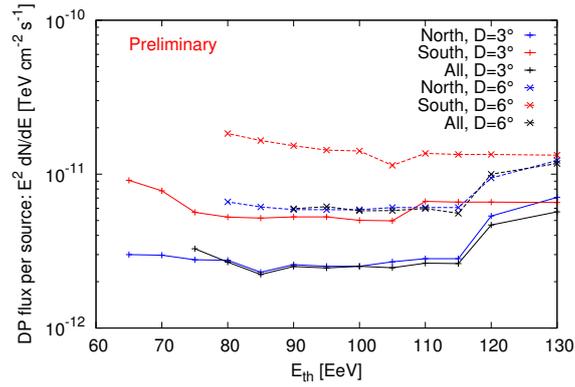

**Figure 5:** Flux normalization required per source for a discovery potential (DP) of $5\sigma$ (pre-trial) for the stacking analysis with the neutrino point-source data set and UHECRs with energies above values of $E_{\text{th}}$, for the Northern and Southern skies separately and together, for the assumed deflections $\sigma_{\text{MD}}(E_{\text{CR}}) = D \times 100\,\text{EeV}/E_{\text{CR}}$, $D = 3°, 6°$.